\documentclass[12pt,catalan]{article}
\usepackage{graphicx}
\usepackage{amsmath}
\usepackage{amsfonts}
\usepackage{amssymb}
\usepackage{bm}
\usepackage{cite}

\usepackage[english]{babel}
\usepackage[latin1]{inputenc}
\usepackage{times}
\usepackage[T1]{fontenc}

\newcommand{\beq}{\begin{equation}}
\newcommand{\eeq}{\end{equation}}
\newcommand{\bea}{\begin{eqnarray}}
\newcommand{\eea}{\end{eqnarray}}

\newcommand{\RR}{\ensuremath{ \mathbb{R}} }

\newcommand{\CC}{\ensuremath{ \mathbb{C}} }

\newcommand{\NN}{\ensuremath{ \mathbb{N}} }

\newcommand{\D}{{\rm d}}

\def\bea*{\begin{eqnarray*}}
\def\eea*{\end{eqnarray*}}

\newcommand{\realp}{{\mathbb{R}}{\rm e}}
\newcommand{\tg}{{\rm tg\,}}

\usepackage{tikz}
\usetikzlibrary{arrows}
\tikzstyle{block}=[draw opacity=0.7,line width=1.4cm]

\begin{document}
\title{Infinite Derivatives vs Integral Operators. \\
The Moeller-Zwiebach Puzzle}
\author{Carlos Heredia\thanks{e-mail address: carlosherediapimienta@gmail.com}\,\,\, and Josep Llosa\thanks{e-mail address: pitu.llosa@ub.edu}\\
Facultat de F\'{\i}sica (FQA and ICC) \\ Universitat de Barcelona, Diagonal 645, 08028 Barcelona, Catalonia, Spain }
\maketitle

\begin{abstract}
We study the relationship between integral and infinite-derivative operators. In particular, we examine the operator $p^{\frac12\,\partial_t^2}\,$ that appears in the theory of $p$-adic string fields,  as well as the Moyal product that arises in non-commutative theories. We also attempt to clarify the apparent paradox presented by Moeller and Zwiebach, which highlights the discrepancy between them.
\\[1ex]
\noindent
\end{abstract}

\section{Introduction: A short chronological review \label{S1}}
Theoretical physics has resorted to nonlocal Lagrangians and infinite-order Lagrangians in several instances. As far as we know, a chronological list should include
%\begin{description}
%\item[1.-] 
Fokker-Wheeler-Feynman electrodynamics \cite{Fokker1929,Feynman1949}, an attempt to set up electrodynamics without an intermediate field. The Lagrangian has two contributions, namely a part for free point charges $L_0$ plus an interaction term $L_{\rm int}$ resulting from the addition of the symmetric Li\'enard-Wiechert potentials ---half-advanced and half-retarded--- of each pair of charges. 
%\item[2.-] 

Then, in the fifties, there were several attempts, such as that of Pais and Uhlenbeck  \cite{Pais1950}, who proposed higher-order field equations ---even equations of infinite order--- ``to eliminate the divergent features of the present theory'', or the proposals by Yukawa \cite{Yukawa1950} or Kristensen and M\o ller \cite{Kristensen1952} to avoid divergent perturbation expansions in nucleon-meson interaction. 

Later, in the seventies, we find Marnelius' significant contribution \cite{Marnelius1973} to formulating a variational principle and conserved currents for nonlocal field theories. He starts with an action made of a local free part plus a nonlocal interaction part
\begin{equation}  \label{i0}
 S_{\rm int}(\phi) = \int \D x_1 \ldots \D x_N \,K_{AB\ldots D}(x_1, \ldots x_N) \phi^A(x_1) \ldots \phi^D(x_N)  \,,
\end{equation}
i.e., already shaped in the ``integral operator'' (IO) fashion, which he turns into an infinite derivative (ID) form by means of the {\em formal} Taylor expansion
\begin{equation}  \label{i1}
\phi(x) = \sum_{|\alpha|=0}^\infty \frac{(x-X)^\alpha}{\alpha!}\,\partial_\alpha \phi(X)\,,
\end{equation}
where we have used the multi-index notation, namely, $\,x \in \RR^n \,,\,\,\,(\alpha_1, \ldots,\alpha_n)\in\mathbb{N}^n_0\,$,
$$ \partial_\alpha := \partial_1^{\alpha_1} \ldots \partial_n^{\alpha_n}\,, \quad \quad  x^\alpha := (x^1)^{\alpha_1} \ldots (x^n)^{\alpha_n}\,, \quad \quad \alpha!:= \alpha_1! \ldots\alpha_n!  $$
and $\, |\alpha| := \alpha_1 +\ldots +\alpha_n \,$. %$$\alpha = (\alpha_1, \ldots , \alpha_k)  \qquad{\rm and} \qquad |\alpha| = k \,, \qquad \alpha_j = 1 \ldots 4 \,. $$ 
This procedure transforms the action into
\begin{equation}  \label{i0a}
 S_{\rm int}(\phi) = \int \D X\, \sum_{|\alpha|,\ldots |\gamma|=0}^\infty E_{AB\ldots D}^{\alpha \beta \ldots \gamma}(X)\,\partial_\alpha\phi^A(X) \ldots \partial_\gamma\phi^D(X) \,, 
\end{equation}
which is (pseudo)local in that it contains field derivatives of any order.

Once the action is written in this form, the variational principle and Noether's theorem can be applied in the standard way for a Lagrangian of order $n$ ---but replacing $n$ with $\infty$--- to derive: (i) the field equations whose interaction term depends on the field derivatives of any order, and (ii) the conserved current associated to any continuous symmetry of the Lagrangian (again the interaction contribution contains a series with all field derivatives). Finally, Marnelius manages to sum those ID series and write the interaction terms, both in the field equations and conserved currents, in the ``integral operator'' shape.

The method relies on the formal Taylor expansion (\ref{i1}), which is abusive and leads to a glaring inconsistency. Indeed, not all functions on which the action is defined are real analytic functions with an infinite radius of convergence, not even if we limit ourselves to the solutions of the field equations. The inconsistency is that, whereas in finite order ($n$) Lagrangians, the variational principle reads
$$ \delta S = 0 \qquad{\rm for} \qquad \delta \partial_\alpha \phi(B) = 0 \,, \qquad |\alpha| \leq n-1 \,,$$
$B$ meaning the boundary. When we replace $n$ with $\infty$, it becomes
$$ \delta S = 0 \qquad{\rm for} \qquad \delta \partial_\alpha \phi(B) = 0 \quad\mbox{for all } \alpha=(\alpha_1, \alpha_2, \ldots )\,, $$
which, due to the Taylor expansion assumption, implies that
$$ \delta S = 0 \qquad{\rm for} \qquad \delta \phi(x) = 0 \quad\mbox{for all } x  \,,  $$
which is a triviality.

One may reply to this objection by saying that Marnelius' method is merely an intermediate heuristic trick which, despite its inconsistencies, allows to derive useful field equations and conserved currents, whose local conservation on-shell can be checked a posteriori. In our view, this fact indicates that there must be a consistent reasoning leading to the same results.

%\item[3.-]  
In the eighties, nonlocality arises again as ``\ldots a fundamental property of string field theory and not just as an artifact \ldots'' \cite{ELIEZER1989}. They deal with it in the ``higher derivative representation'' (in their terminology), consisting in truncating the infinite derivative series to work it as a higher (finite) derivative theory and then replacing $n$ with infinity.

%\item[4.-] 
More recently, nonlocality --either as IO or ID-- appears in modified gravity theories. Inspired by string theory's ultraviolet (UV) finiteness, they are being proposed to solve both cosmological and black hole singularities (see, for instance, \cite{Biswas2006,Modesto2012,Biswas2012,Frolov2016_2} and references therein). Likewise, they are also being used to explain the cosmic expansion of the Universe without relying on a contribution from dark energy \cite{Deser2007}, inflationary models \cite{Briscese2013}, and gravitational waves \cite{Calcagni2022_2}.  
%\end{description}

The central question is: are both approaches, ID and IO, equivalent? In what follows, we will examine the operator $p^{\frac12\,\partial_t^2}$ that appears in the $p$-adic particle, i.e the $p$-adic string field independent of spatial coordinates \cite{BREKKE1988, Vladimirov2004}, as well as the Moyal product, which arises in non-commutative theories \cite{Seiberg1999, Gomis2001_2}.

\section{The operator $p^{\frac12\,\partial_t^2}$}

The expression of the operator $p^{\frac12\,\partial_t^2}\,$ in the two different approaches is:
%\begin{description}    \item[ID]{ 
\begin{equation}  \label{1.1a}
\mathbf{ID} \hspace*{5em} \hat K_1 \psi(t) := \sum_{n=0}^\infty \frac{r^n}{n!}\,\psi^{(2n)}(t) \,, \qquad r = \frac12\,\log p > 0 \,,
\end{equation} 
where $p$ is a prime integer, and   
%}    % \item[IO]{ 
\begin{equation}  \label{1.1b}
\mathbf{IO} \hspace*{5em} \hat K_2 \psi(t) := \left(\mathcal{K}_r \ast \psi \right)_{(t)} \,, \qquad {\rm with} \qquad \mathcal{K}_r(t) = \frac1{2\sqrt{\pi r}}\,e^{\frac{-t^2}{4 r} }\,.
\end{equation}
%}    \end{description}
Rescaling the variable $\tau= t/\sqrt{r}\,$, both operators keep the same form replacing the parameter $r$ with 1. Hereon, we will analyze the case $r=1$ because it presents the same features as the general case. Furthermore, it is worth mentioning that the connection between operators (\ref{1.1a}) and (\ref{1.1b}) is derived in Section \ref{S1.1} below, and it relies on the use of Fourier transforms \cite{ELIEZER1989,Carlsson2016_E4}.

In mathematical literature, representation (\ref{1.1b}) is known as the Weierstrass transform, which was extensively studied around the 1960s. For instance, the necessary and sufficient conditions for this representation to be well-defined were discussed in ref.\cite{widder1951}. Additionally, the inverse of this transformation, denoted as $e^{\partial^2_t}$, was examined in ref. \cite{Sumner1957}, shedding light on the ineffectiveness of replacing the inverse with a Taylor series, given that many functions diverge under such an approach\footnote{For readers interested in further extensions of these studies, ref. \cite{zemanian1967} presents a generalization of this transform and its inverse for generalized functions.}. It should also be mentioned that Hermite polynomials have been used to study the interpretability and summability of the inverse Weierstrass transform when interpreted as infinite series \cite{Bilodeau1962}. For more recent literature, it is worth mentioning alternative derivations \cite{Calcagni2008, Kolar2022} that are based on the heat kernel.

\subsection{The puzzle \label{S1.0}} 
Moeller and Zwiebach \cite{Moeller2002} pointed out that $\hat K_1 \psi$ and $\hat K_2 \psi$ are actually different for a smooth, compact support function such that
\begin{equation}  \label{1.1c}
 \psi(t) = 1\,,\qquad{\rm if} \qquad |t|\leq a \,, \qquad {\rm and}  \qquad \psi(t) = 0\,,\qquad{\rm if} \qquad |t|\geq b \,,
\end{equation} 
because 
$$ \hat K_1 \psi(0) = 1\,,\qquad{\rm if} \qquad |t|< a\,,\qquad {\rm and}  \qquad \hat K_1 \psi(t) = 0\,,\qquad{\rm if} \qquad |t|>b \,,$$
whereas
$$ \hat K_2 \psi(0) < 1\,,\qquad{\rm for} \qquad |t|< a\,,\qquad {\rm and}  \qquad \hat K_2 \psi(t) > 0\,,\qquad{\rm for} \qquad |t|>b \,.$$

This said, the following questions arise: (a) Which operator, $\hat K_1 \psi$ or $\hat K_2 \psi$, is ``the good one''? (b) do they act on the same domain of functions?, and (c) is there any subdomain in which both operators coincide? 

%To shed some light and make things more clear, the following considerations are relevant. 
We must always bear in mind that the operators $\hat K_1 \psi$ and $\hat K_2 \psi$ usually occur in a Lagrangian (and an action integral), and therefore, for $\hat K_a$ to be a ``valid'' operator, $\hat K_a \psi(t)$ must be well-defined for any $t\in \RR\,$, in some sense. 

As for the domains of definition, it is evident that $\hat K_1$ acts on smooth functions $ \psi \in\mathcal{C}^\infty(\RR)\,$, and moreover, the series defining $\hat K_1 \psi$ must be ``summable'' in some sense, namely pointwise, uniformly, or the like\footnote{Although the requirement of pointwise convergence might be loosened and replaced with {\em convergence almost everywhere}, this seems more appropriate for a mathematical context rather than a mechanical one.}. 

\subsection{How are the operators $\hat K_1$ and $\hat K_2$ connected? \label{S1.1}}
As pointed out in ref. \cite{ELIEZER1989}, provided that $\psi(t)$ has a Fourier transform 
$$ \mathcal{F}\psi (k) = \tilde\psi(k) = \frac1{\sqrt{2\pi}}\,\int_\RR \D t\,e^{ikt}\,\psi(t) \qquad {\rm and} \qquad 
\overline{\mathcal{F}}\tilde\psi (t) = \psi(t) = \frac1{\sqrt{2\pi}}\,\int_\RR \D k\,e^{-ikt}\,\tilde\psi(k) \,, $$
both operators are connected through the following steps:
\begin{description}
\item[$(1^{\rm st})$] Substituting and differentiating under the integral sign, we have that
\begin{equation}  \label{1.2a}  
\hat K_1 \psi(t) = \frac1{\sqrt{2\pi}}\,\sum_{n=0}^\infty \frac1{n!}\,\frac{\D^{2n}\;}{\D t^{2n}}\int_\RR \D k\,e^{-ikt}\,\tilde\psi(k) = \frac1{\sqrt{2\pi}}\,\sum_{n=0}^\infty \frac1{n!}\,\int_\RR \D k\,e^{-ikt}\,(- k^2)^n\,\tilde\psi(k)  \,.
\end{equation}
Then, 
\item[$(2^{\rm nd})$] commuting the series and the integral signs, we can write
\begin{equation}  \label{1.2b}
 \hat K_1 \psi(t) =  \frac1{\sqrt{2\pi}}\,\int_\RR \D k\,e^{-ikt}\,\sum_{n=0}^\infty \frac{(- k^2)^n}{n!}\,\tilde\psi(k) = \frac1{\sqrt{2\pi}}\,\int_\RR \D k\,e^{-ikt- k^2}\,\tilde\psi(k)
\end{equation}
and, since $\,\displaystyle{ \frac1{\sqrt{2\pi}}\,\int_\RR \D k\,e^{-ikt- k^2} =  \frac1{\sqrt{2 r}}\,e^{\frac{-t^2}{4}} }\,$, 
\item[$(3^{\rm rd})$] by the convolution theorem, we arrive at
$$   \hat K_1 \psi(t) = \left(\mathcal{K}_1 \ast \psi \right)_{(t)} = \hat K_2 \psi(t) \,.   $$
\end{description}

The soundness of the above chain of reasoning relies on the validity of the three highlighted steps: ($1^{\rm st}$) differentiation under the integral sign, ($2^{\rm nd}$) the commutability of the series and the integral, and ($3^{\rm rd}$) the convolution theorem. Some mathematical theorems provide sufficient conditions for each one to be allowed. While the conditions for steps $1^{\rm st}$ and $3^{\rm rd}$ to be valid are reasonably mild, the validity of step $ 2^{\rm nd}$ might rely either on the termwise integration theorem for uniformly convergent series \cite{Apostol1976_TH913} or on Lebesgue's dominated convergence theorem \cite{Apostol1976_V2}, namely\\[1.5ex]
{\bf Theorem [Lebesgue]}\\[1ex]
{\em Let $\,\{f_n\}\,$ be a sequence of  Lebesgue integrable functions in an interval $\mathcal{I}\,$. Assume that: (i) the sequence  converges to $\,f\,$ pointwise almost everywhere (a.e.w.) in $\mathcal{I}\,$, and (ii) it exists a non-negative Lebesgue integrable function $\,g \,$ such that 
$$ |f_n(z)| \leq g(z) \,, \qquad \mbox{a.e.w. in }\;\mathcal{I} \,, \qquad \forall n \,. $$
Then $f$ is  Lebesgue integrable and $\; \displaystyle{\int_\mathcal{I} f(z)\,\D z = \lim_{n\rightarrow\infty} \int_\mathcal{I} f_n(z)\,\D z}\,$.}\\[2ex]
(In our case, $\,\{f_n(z)\}\,$ is the sequence of partial sums of a series.)

The present work aims to address the apparent paradox raised in ref. \cite{Moeller2002} and to understand why the action of the operator $\,e^{r\partial_t^2}\,$ on a bounded support function yields a different result, depending on whether the ID form $\hat{K}_1$ or the IO form  $\hat{K}_2$ is used.

We will prove that $\hat{K}_1 \eta $ is not globally defined for the particular finite smooth function $\eta(t)\,$ considered in ref. \cite{Moeller2002}, i.e., the series diverges for a wide range of values of $t\,$.

Carlson et al. \cite{Carlsson2016_E4} did already prove that, when acting on a finite smooth function, a certain class of operators, which are defined by an infinite series containing derivatives of any order, lead to divergent series, and therefore they are not globally defined. The operator $\,e^{r\partial_t^2}\,$ belongs to this class. However, instead of invoking that result in ref. \cite{Carlsson2016_E4}, which is based on a rather powerful and sophisticated mathematical formalism, we think that our contribution ---however clumsy it may be--- might be helpful in that, on the one hand, it is more accessible to a wider community and, on the other, we emphasize pointwise convergence that, in our view, is more suitable than $L^1$ convergence for a classical physics context.   

\subsection{The domains of the operators $\hat K_1$ and $\hat K_2$   \label{S2}}
There is a common domain where both operators coincide, but apparently, the domain of $\hat K_2$ is wider than that of $\hat K_1$, as illustrated by the following examples.

\subsubsection{Entire functions \label{S2.1}}
$\hat K_2$ and $\hat K_1$ coincide on the space of entire functions, i.e., real functions that result from restricting entire complex functions to $\RR$. Indeed, let
\begin{equation} \label{e2.1}
\psi(t) = \sum_{l=0}^\infty \frac{\psi^{(l)}(t_0)}{l!}\,(t-t_0)^l  \,, \qquad\forall t\,, \;t_0 \in \RR \,.
\end{equation}
Applying the operator $\hat K_2$, we have
\begin{eqnarray}
 \hat K_2 \psi(t) &=&\frac1{2\sqrt{\pi }}\,\int_\RR\D\tau\, e^{\frac{-\tau^2}{4}} \psi(t-\tau) =
\frac1{2 \sqrt{\pi}}\,\int_0^\infty \D\tau\, e^{\frac{-\tau^2}{4}} \,\left[\psi(t-\tau) + \psi(t+\tau)\right]  \nonumber\\[2ex]
\label{e2.2}
    & = & \frac1{\sqrt{\pi}}\,\int_0^\infty \D\tau\,\sum_{n=0}^\infty \frac{\psi^{(2n)}(t)}{(2n)!}\,\tau^{2n} \, e^{\frac{-\tau^2}{4}} \,,
\end{eqnarray}
and since 
$$\sum_{n=0}^\infty {e^{\frac{-\tau^2}{4}}}\,\frac{\psi^{(2n)}(t)}{(2n)!}\,\tau^{2n} = 
\frac12\,{e^{\frac{-\tau^2}{4}}}\,\left[\psi(t-\tau) + \psi(t+\tau)\right] \,$$
uniformly on any interval  $\,[-K,K] \,,\quad \forall K\in \RR^+\, $, the series may be termwise integrated \cite{Apostol1976_V2}.

The individual integrals are easily performed
$$ \int_0^\infty \D\tau\,\tau^{2n} \, e^{\frac{-\tau^2}{4}} = \frac{(2n)! \sqrt{\pi}}{n!}\,,$$
and by substituting them in (\ref{e2.2}), we finally obtain 
\begin{equation} \label{e2.3}
 \hat K_2 \psi(t) = \frac1{\sqrt{\pi}}\,\sum_{n=0}^\infty \frac{\psi^{(2n)}(t)}{(2n)!}\,\frac{(2n)! \sqrt{\pi}}{n!} = 
\sum_{n=0}^\infty \frac{1}{n!}\,\psi^{(2n)}(t) = \hat K_1 \psi(t)\,. 
\end{equation}

\subsubsection{An apparently ``harmless'' function \label{S2.2}}
Let us consider the function 
\begin{equation} \label{e2.3a}
 \psi(t) = \displaystyle{\frac1{1+t^2} }\,.
\end{equation}
Although it is $\mathcal{C}^\infty(\RR)$, bounded, and also $L^p(\RR)\,$, its analytic extension does not yield an entire function because it presents two single poles at $t=\pm i\,$.

It is a familiar function because, apart from normalization, it is Cauchy probability density and its Fourier transform is  
$$\tilde\psi(k) = \sqrt{\frac\pi2}\, e^{-|k|}\,. $$ 

It belongs to the domain of $\hat K_2$; indeed, 
\begin{equation} \label{e2.4}
  \hat K_2 \psi(t) = \frac1{2\sqrt{\pi}}\,\int_\RR \D\tau \, e^{\frac{-\tau^2}4} \, \frac1{1+(t-\tau)^2} 
\end{equation}
is well-defined and finite for any $t\in \RR\,$.

However, despite $\psi$ being so little pathological, it presents serious difficulties for the operator  $\hat K_1$.
We must first obtain all derivatives of even order and, to make it simpler, we include that
\begin{equation} \label{e2.5}
\psi(t) = \frac12\,\left[\psi_+(t) + \psi_-(t)\right] \,, \qquad {\rm with} \qquad \psi_\pm(t) = (1 \pm i t)^{-1} \,,
\end{equation}
whence it quickly follows that $\,\displaystyle{ \psi_\pm^{(2n)}(t) = (2n)! (-1)^n \,(1 \pm i t)^{-(2n+1)}  }\,$. 
Thus, from (\ref{e2.5}), we have that
$$ \psi^{(2n)}(t) = \frac{(2n)! (-1)^n}2 \,\left[(1 + i t)^{-(2n+1)} + (1 - i t)^{-(2n+1)}\right]  \,, $$
which, using the new variable
\begin{equation} \label{e2.5a}
 e^{\pm i\alpha} = \frac{1 \pm i t}{\sqrt{1+t^2}} \,, \qquad \qquad t = \tg \alpha \,,  \qquad |\alpha| < \frac\pi2 \,,
\end{equation}
can be written as
\begin{equation} \label{e2.6}
\psi^{(2n)}(t) = \frac{(2n)! (-1)^n}{(1+t^2)^{n+1/2}}\,\cos\left[(2n+1) \alpha\right] \,
\end{equation}
and, therefore,
\begin{equation} \label{e2.7}
 \hat K_1 \psi(t) = \sum_{n=0}^\infty a_n(t) \,, \qquad {\rm with} \qquad
a_n(t) := \frac{(2n)! (-1)^n}{n!\,(1+t^2)^{n+1/2}}\,\cos\left[(2n+1) \alpha\right]  \,.  
\end{equation}

We will prove that the series is divergent for all $t\in \RR\,$. Our proof relies on a corollary of Cauchy's condition \cite{Apostol1976_V3} that provides a necessary condition for pointwise convergence, namely $ \,\displaystyle{\lim_{n\rightarrow\infty}|a_n(t)| = 0 }\,,$
and therefore the sequence 
$$ \quad A := \{|a_n(t)|\} \quad \mbox{must be bounded for the series to converge.} $$

It is obvious that for $t=0$ and $\alpha = 0$, the sequence $\{|a_n(0)|\}$ is unbounded. For other values of $t$, we restrict to positive ones and distinguish two cases:
\begin{description}
\item[If $\alpha/\pi$ is rational,] then $\alpha/\pi $ is an irreducible fraction $\, k/l < 1/2\,$ and  
$$ (2n+1)\alpha = D_n \pi + \frac{s_n}{l}\,\pi \,, \qquad D_n, s_n\in \NN\,, \quad 0 \leq s_n < l \,,$$
where $D_n$ and $s_n < l$ denote the integer quotient and remainder, respectively. Therefore,
$$ \left|\cos(2n+1)\alpha \right| = \left|\cos\left(\frac{s_n \pi}{l}\right) \right|\,,$$
and, as $n$ runs over $\NN$, it repeats periodically, yielding a finite set of positive numbers (some may be null but, as a rule, not all of them vanish), namely
$$  \left\{\left|\cos(2n+1)\alpha \right| \,, \quad n \in \NN \right\} = \{ p_0 < p_1 < \ldots < p_n\} \,.$$

Now, pick a $p \neq 0$ among these and consider the subsequence $F_p \subset A\,$ corresponding to those $m$ such that 
$\,\displaystyle{\left|\cos(2m+1)\alpha \right|= p  }\,$, namely
$$ F_p = \left\{\frac{(2m)!}{m!(1+t^2)^{m+1/2}}\,p     \,, \quad m \in \NN \,, \; \left|\cos(2n+1)\alpha \right|= p \right\} \,.$$
This sequence is not bounded as it easily follows from Stirling's formula
$$ \frac{(2m)!}{m!(1+t^2)^{m+1/2}} \sim \frac{\sqrt{2}}{\sqrt{1+t^2}}\,\left(\frac{4 m}{e (1+t^2)} \right)^m  \rightarrow \infty \,.$$
Therefore, $A$ is unbounded, and the series (\ref{e2.7}) is divergent for those $t = \tg \alpha \,$, where $\alpha$ is a rational multiple of $\pi\,$.

\item[If $\alpha/\pi$ is irrational,] then $e^{i(2n+1)\alpha}$ does not repeat periodically as a function of $n$, and the sequence 
$$ E:= \left\{ e^{i(2n+1)\alpha}, \; n\in \NN\right\} \subset \CC \qquad\mbox{is infinite and bounded.} $$
Then, the Bolzano-Weierstrass theorem \cite{Apostol1976_V4} implies that $E$ has at least an accumulation point different from $\pm i\,$. 
Indeed, assume that $i$ was the only accumulation point, then there would exist a subsequence, $\, E_1:= \left\{ e^{i(2n_j+1)\alpha} , \; j\in \NN\right\}  \,$,  
such that
$$\lim_{ j \rightarrow \infty} e^{i(2n_j+1)\alpha} = i  \,. $$
Consider now the subsequence resulting from choosing the next of each term in $E_1$, that is, $\, E_2:= \left\{ e^{i(2[n_j+1]+1)\alpha}, \; j\in \NN\right\} \,$. It is obvious that 
$$ \lim_{j \rightarrow \infty}  e^{i(2[n_j+1]+1)\alpha} =   e^{i\,2\alpha}\,\lim_{j \rightarrow \infty} e^{i(2n_j+1)\alpha}  i\, e^{i\,2\alpha}\,,$$
and, as $2\alpha <\pi\,$, we have proved that $\,i e^{i\,2\alpha} \neq\pm i \,$ is an accumulation point of $E$ as well.

For the $n_j +1$ term in the subsequence $E_2$, we have that
$$ \lim_{j\rightarrow\infty} \left|\cos(2[n_j+1]+1)\alpha \right| = 
\left|\cos\left(\frac\pi2 +2 \alpha \right)\right| = 
\left|\sin (2\alpha) \right| \neq 0 \,.$$
Therefore, $\,p=\left|\sin (2\alpha) \right| \,$ is an accumulation point of $\displaystyle{ \left\{\left|\cos(2n+1)\alpha \right| \,, \quad n \in \NN \right\} }\,$, and, given the positive number $p/2$, there exist an infinite number of values of $n$ for which 
$$ \left|\cos(2 n +1)\alpha \right| > \frac{p}2 \,. $$
If we then consider the subsequence $F_1 \subset A$ corresponding to those $m$ such that \\
$\,\displaystyle{\left|\cos(2m+1)\alpha \right|> p/2  }\,$, we quickly see that $F_1$ is unbounded. Therefore, the series (\ref{e2.7}) is also divergent for those $t = \tg \alpha \,$, such that $\alpha$ is an irrational multiple of $\pi\,$. \hfill $\Box$
\end{description}

\subsubsection{Can Lebesgue's dominated convergence theorem be applied to convert $\hat K_1$ into $\hat K_2$ for this function?   \label{S2.2.1}}
A consequence of the results so far is that Lebesgue's theorem does not apply in the case of the function (\ref{e2.3a}). Indeed, whereas $\hat{K}_2 \psi$ is well defined, $\hat{K}_1 \psi$ runs into an everywhere divergent series. Let us examine in detail how the hypothesis of the theorem does fail.

For this particular function, the conversion of $\hat{K}_1\psi$ into $\hat{K}_2\psi$ requires swapping the integral and the series in (\ref{1.2b}), which is easily written as 
$$ \hat K_1 \psi(t) =  \int_0^\infty \D k\, \sum_{l=0}^\infty \frac{(- k^2)^l}{l!}\,e^{-k}\,\cos(kt) \,. $$
For Lebesgue's theorem to apply, we need that a non-negative dominant function $g \in L([0,\infty[)$ exists such that 
$$ |f_n(k)| \leq g(k) \,, \qquad {\rm with} \qquad f_n(k) = \realp\left(\sum_{l=0}^n \frac{(- k^2)^l}{l!}\,e^{-k(1-it)}\right) \,, \quad \forall\, n \,,$$ 
and therefore 
$$ \left|\int_0^\infty f_n \right| \leq \int_0^\infty \left| f_n \right| \leq \int_0^\infty g\,, \quad \forall\, n \,.$$
Thus, for a dominant function $g$ to exist, the quantities $\int_0^\infty f_n $ must be bounded.
 
Let us now introduce
$$ I_n(\alpha):=\int_0^\infty f_n = \sum_{l=0}^n \frac{(- 1)^l}{l!} \, \realp\int_0^\infty \D k\, k^{2l}\,e^{-k(1-it)} =  \sum_{l=0}^n \frac{(- 1)^l}{l!} \, \realp\left[ \frac{(2 l)!}{(1+ i t)^{2l+1}}\right] $$
that, using the variable (\ref{e2.5}), becomes
$$  I_n(\alpha)= \sum_{l=0}^n \frac{(- 1)^l \,(2l)!}{l! (1+t^2)^{l+1/2}} \, \cos{(2 l+1)\alpha} \,.$$ 

Now, the sequence $\,\{ I_n(\alpha)\,, \; n\in \NN\}\,$ is unbounded. Were it bounded, so would be the sequence $\, \{ I_n(\alpha) -  I_{n-1}(\alpha)\,, \; n\in \NN\}\,$, but 
$$ I_n(\alpha) -  I_{n-1}(\alpha) = \frac{(- 1)^n \,(2n)!}{n! (1+t^2)^{n+1/2}} \, \cos{(2 n+1)\alpha} \,  $$
that, as proved before, is unbounded and, therefore, the dominant function $g$ does not exist.

In summarizing, a simple smooth function as (\ref{e2.3a}), which is square summable but not an entire function, belongs to the domain of $\hat{K}_2$ but not to that of $\hat{K}_1$. Therefore, the $2^{\rm nd}$ step in the conversion of $\hat{K}_1$ into $\hat{K}_2$ must fail, and we have shown why the dominated convergence theorem does not apply in this case.

\subsection{The anomalous function of Moeller and Zwiebach  \label{S2.3} }
We will now study the action of the ID operator $\hat{K}_1$ on the particular finite smooth function considered in ref. \cite{Moeller2002}.  We will see that it is not globally defined, i.e., it fails for a wide range of values of the variable. 

We consider the finite smooth function defined by \cite{Vladimirov_GF_v2}
\begin{equation}   \label{1.3b}
  \eta(t) :=  \int_\RR \D\tau\,\chi(\tau)\,\omega(t-\tau)  = \int_{-c}^c \D\tau\,\omega(t-\tau) \,,
\end{equation}
with 
\begin{equation}   \label{1.3}
  \omega(t) = C\cdot e^{-\frac1{1-t^2}}\,, \qquad |t| < 1 \quad {\rm and}\qquad \omega(t) = 0\,, \quad{\rm otherwise} \,,
\end{equation} 
where $\,\displaystyle{C^{-1} := \int_\RR  e^{-\frac1{1-t^2}} \,\D t}\,$ and 
\begin{equation}   \label{1.3a}
  \chi(t) = 1\,, \quad {\rm if }\quad |t| < c \quad {\rm and}\qquad \chi(t) = 0\,, \quad{\rm otherwise} \,, \qquad {\rm with} \quad c > 1\,.
\end{equation}
$\omega$ is a smooth finite function, i. e. its support $[-1,1]\,$ is bounded. Notice that the parity of its derivatives is
$$ \omega^{(n)}(-t) = (-1)^n\, \omega^{(n)}(t)  \,.$$

Since $\omega(t)$ is even, so is $\eta(t)$, and it suffices to study $\eta(t)$ for $t\geq 0\,$. Thus, including that $\omega(\tau)$ vanishes for $|\tau|> 1$ and that $\,t+c \geq c >1\,$, we have that
\begin{equation}   \label{1.3z}
 \eta(t) = \int_{|t-c| < c} \D\tau\,\omega(\tau) = \int_{t-c}^1 \D\tau\,\omega(\tau) \,,
\end{equation}
whence it follows that
\begin{equation}   \label{1.3c}
 \eta(t) = \left\{ \begin{array}{lcl}
								1 \,, &\quad & {\rm if}\;\; 0 \leq t \leq c-1 \\[1ex]
								0 \,, &\quad &  {\rm if}\;\; t \geq c+1  \\[1ex] 
								\displaystyle{ \eta(t) = \frac12 -  \int_0^{|t-c|} \D\tau\,\omega(\tau) \,}\,, &\quad & {\rm if}\;\; |c-t| < 1\,,
													\end{array}   \right.
\end{equation}
where the facts that $\omega(\tau) $ is even and $\int_{-1}^{+1} \D\tau\,\omega(\tau) = 1 \,$ have been included.

\begin{center}
\includegraphics[width=12cm]{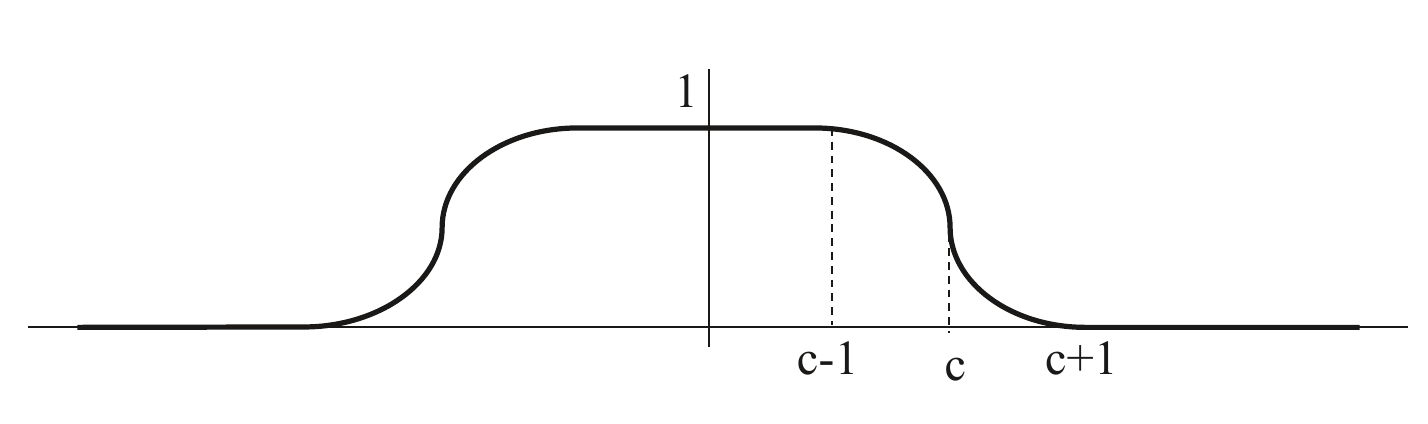}
\end{center}

\noindent 
It is evident that $\eta(t)$ is a particular case of the functions considered in ref. \cite{Moeller2002}. The action of the operator $\hat{K}_2\,$ yields
$$ \hat{K}_2\eta(t) = \int_\RR \D\tau\,\int_{-c}^c \D\tau^\prime\,K_r(t-\tau)\,\omega(\tau-\tau^\prime) \,,$$
which is well-defined for any finite smooth $\omega \,$.

To study the action of $\hat{K}_2\,$, we need the derivatives of $\eta$ for $n \geq 1\,$, which are better obtained directly from (\ref{1.3b}), and we have
$$ \eta^{(n)}(t) = \int_{-c}^c \D\tau\,\omega^{(n)}(t-\tau) = (-1)^n\,\int_{-c}^c \D\tau\,\omega^{(n)}(\tau-t) \,.$$
For $t$ positive, $c+t >1\,$ and $\omega(c+t)=0\,$. Therefore,
\begin{equation}   \label{1.3d}
\eta^{(n)}(t) = (-1)^n \,\omega^{(n-1)}(c-t) \,,\qquad n\geq 1 \,
\end{equation}
or
\begin{equation} \label{1.3f}
 \eta^{(n)}(t) = \left\{ \begin{array}{lcl}
													0\,, &\quad & {\rm if}\;\; |c-t| \geq 1 \\[1ex]
								(-1)^n \,\omega^{(n-1)}(c-t) \,, &\quad &  {\rm if}\;\;|c-t| \leq 1  \,,
													\end{array}   \right.
\end{equation}
where we have included that $ c+t >1$, and therefore $\omega^{(n-1)}(c+t) = 0\,$.
Some particular values are
\begin{equation} \label{1.3e}
 \eta^{(n)}(0) = 0 \,, \quad \quad \eta^{(n)}(c\pm 1) = (-1)^n \omega^{(n-1)}(1) = 0
 \,, \quad \quad   \eta^{(n)}(c) =  (-1)^n \omega^{(n-1)}(0) \,.
\end{equation}

Then, using (\ref{1.3f}) for even values of $n$, we obtain that 
$$ \hat K_1\eta(t) = \eta(t) + \sum_{l=1}^\infty \frac1{l!}\,\omega^{(2l-1)}(c-t) $$
and, introducing the variable $\,\tau = c-t\,$, we have
\begin{equation}   \label{1.3j}
\hat K_1\eta(c-\tau) = \eta(c-\tau) + \sum_{n=1}^\infty \frac1{n!}\,\omega^{(2n-1)}(\tau) \,.
\end{equation}
It is obvious that  
$$ \hat K_1\eta(t) = 1 \,,\quad {\rm for\; \;}|t| < c-1 \qquad {\rm and} \qquad  \hat K_1\eta(t) = 0 \,,\quad {\rm for\; \;}|t| > c-1 \,.$$

Thus, we need the derivatives of $\omega(\tau)$ at all odd orders. For practical reasons, we introduce the variable 
$$ x = \frac1{1-\tau^2} \,, \quad \mbox{therefore, \quad  $ 0 \leq \tau < 1 \quad$ amounts to} \quad  \infty > x \geq 1\,, $$
and have that
\begin{equation}  \label{s20}
 \frac{\D x}{\D \tau} = 2 \,\tau\,x^2 \,, \qquad  \omega(\tau) = C\,e^{-x}  \,, \qquad \omega^\prime(\tau) = - 2 \tau x^2 \omega(\tau)   \,.
\end{equation}
We will prove that the derivatives of $\omega(\tau)$ are:
\begin{eqnarray}  \label{P1a}
\mbox{even order} &\qquad & \omega^{(2n)}(\tau) = \omega(\tau)\cdot P_n(x) \, ,\quad \mbox{ for all $n$ and} \\[1ex] \label{P1b}
\mbox{odd order} &\qquad & \omega^{(2n-1)}(\tau) = \omega(\tau)\cdot \tau \cdot Q_n(x) \,, \quad\forall n \geq 1 \,,  
\end{eqnarray}
where $P_n$ and $Q_n$ are polynomials in $x$.
Indeed, from (\ref{s20}), we have that $ \,\omega^\prime(\tau) =  - 2\,\tau \, x^2 \,\omega(\tau)  \, $,
whence the proposition holds for $n=0$ with 
\begin{equation}  \label{s20a}
 P_0=1  \qquad {\rm and} \qquad   Q_1 = - 2 x^2\,.
\end{equation}

Deriving now $\omega^{(2n)}(\tau)$, we have that 
$$ \omega^{(2n+1)}(\tau) = \omega(\tau)\, \tau \,2 x^2\,\left[P^\prime_n(x) - P_n(x)\right]\,,$$
and, therefore,
\begin{equation}  \label{s16}
Q_{n+1}(x) = 2 x^2\,\left[P_n^\prime(x) - P_n(x) \right]\,.
\end{equation}

Similarly, by deriving $\omega^{(2n-1)}(\tau)$, we have that 
$$ \omega^{(2n)}(\tau) = \omega(t)\, \left\{ Q_n(x) \,\left[1 - 2 \tau^2 x^2 \right] + 2 \tau^2 x^2\, Q^\prime_n(x)\right\}\,,$$
and, using that $\;\displaystyle{2 \tau^2 x^2 = 2 x (x-1)  }\,$, we arrive at
\begin{equation}  \label{s16a}
P_n(x) = \left[1+2x-2 x^2\right] \, Q_n(x)+ 2 x [x-1] Q_n^\prime(x)\,. 
\end{equation}
Combining equations (\ref{s16}) and (\ref{s16a}), we finally obtain
\begin{equation}  \label{s16b}
Q_{n+1}(x) = Q_n(x)\,2 x^2\,\left[2 x^2- 6 x + 1\right] + Q^\prime_n(x)\,2 x^2\,[-4 x^2 + 8 x - 1] + 
Q_n^{\prime\prime}(x)\,4 x^3\, [x-1] \,,
\end{equation}
which provides an iterative formula to derive $Q_n(x)$ from  $Q_1(x)\,$. The relations (\ref{P1a}-\ref{P1b}) are followed by the induction principle.

Back to equation (\ref{1.3j}), including (\ref{s20}) and (\ref{P1a}-\ref{P1b}), we arrive at
\begin{equation}   \label{1.3k}
 \hat K_1\eta(c-\tau) = \eta(c-\tau) +  C \,e^{-x}\,\sqrt{\frac{x-1}{x}}\,\sum_{n=1}^\infty \frac{Q_n(x)}{n!} \,;
\end{equation}
therefore, $\,\hat K_1\eta(c-\tau)\,$ is well-defined if, and only if, the series on the right-hand side converges. A necessary condition for convergence is that $\,\displaystyle{ \lim_{n\rightarrow\infty}\frac{\left|Q_n(x)\right|}{n!} = 0}\,$. 

To prove that the series diverges for all $|\tau| < 1\,$, i. e. $x>1\,$, we will show that the general term in the series, $\,\displaystyle{ q_n(x):= \frac{\left|Q_n(x)\right|}{n!}}\,$, is unbounded, but, as the expressions are not so simple as in Section \ref{S2.2}, we have to resort to a numerical simulation with Mathematica\footnote{For the reader who is interested in the code developed for the numerical simulation, please see: \textit{github.com/carlosherediapimienta/IDvsIO}.}.
 
\begin{figure}[ht]  
  \centering
    \includegraphics[width=1\textwidth]{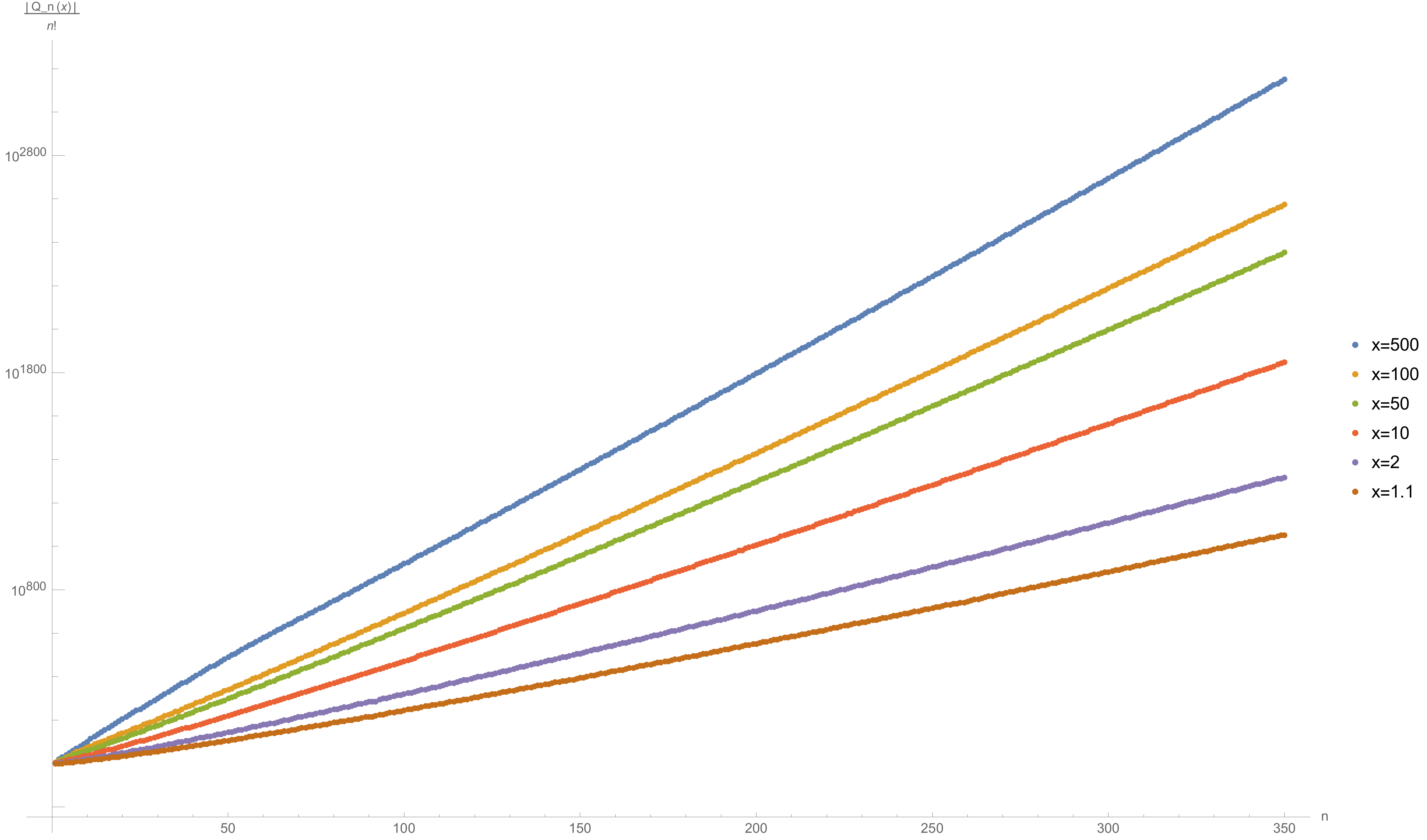}
    \caption{The general term $\,q_n(x)\,$ increases with $n$.\label{graf1}}
\end{figure}

%%%%%%%%%%%%%%%%CARLOS' MODIFICATION
Figure \ref{graf1} is a plot of the general term $\, q_n(x)\,$ for several values of $x$ as a function of $n$ and it shows that the larger is $n$, the more $\, q_n(x)\,$ grows, which clearly indicates that the general term is unbounded. We have tested it using different values of $x=\{1.1, 2, 10, 50, 100, 500\}$ and $n=1,2,\ldots,350$ and found that even for $n = 350$ and $x = 500$, the value of $\,q_n(x)\,$ exceeds $10^{2800}$. We have also tried even larger values of $n$ and $x$, and the result remains the same.

\begin{figure}[ht]   
  \centering
    \includegraphics[width=1\textwidth]{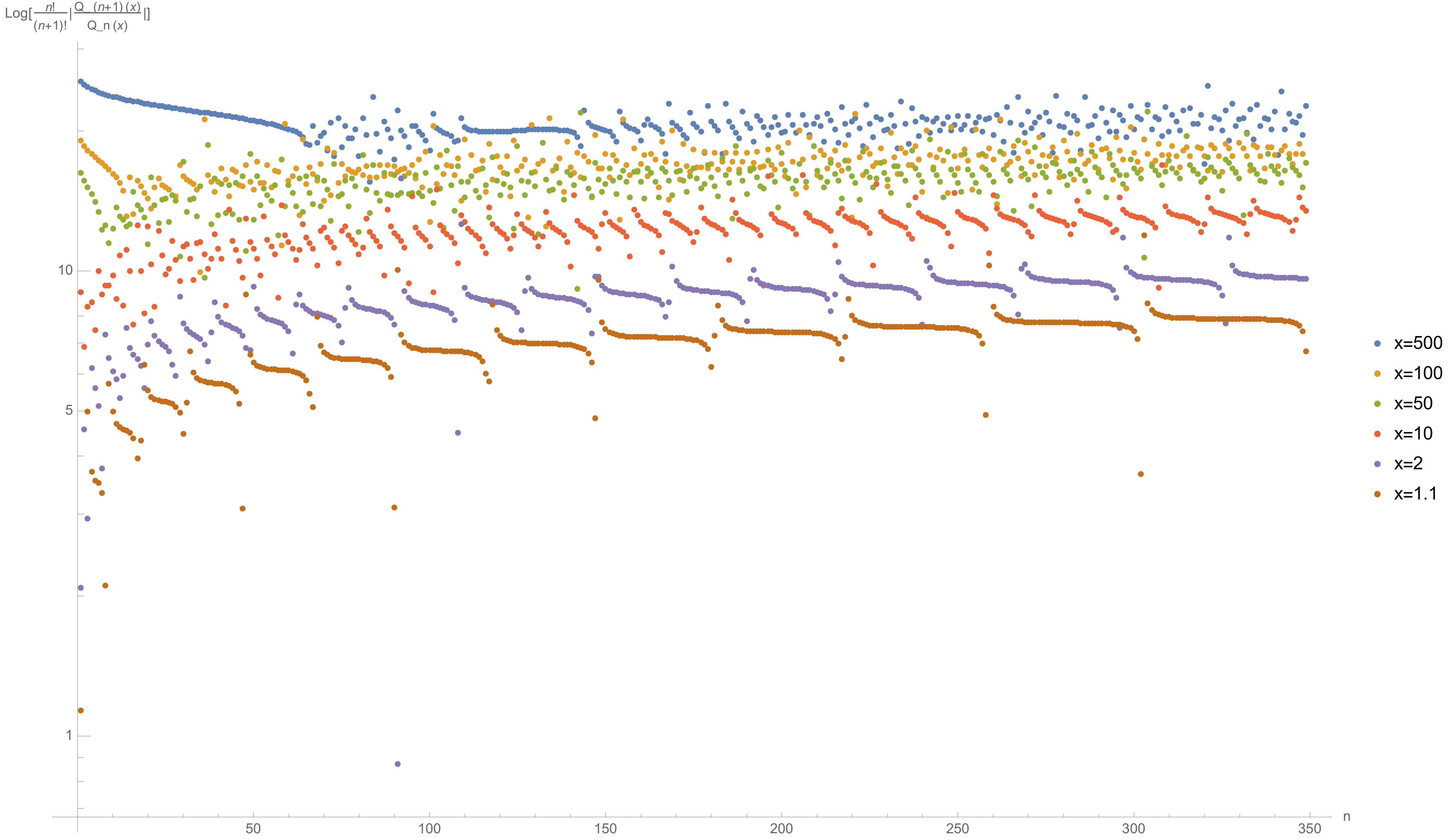}
    \caption{$\log\left[\frac{q_{n+1}(x)}{q_{n}(x)}\right]$ vs $n$.\label{graf2}}
\end{figure}

Figure \ref{graf2} is a logarithmic plot of the quotient $\,\displaystyle{\frac{q_{n+1}(x)}{q_{n}(x)}}\,$.
Notice that the result is strictly positive, indicating that the term $q_{n+1}(x)$ is larger than $q_n(x)$. We have used the same values of $x$ and $n$ as in the previous figure and also tested larger values of $n$, leading to the same conclusion. Therefore, these numerical simulations confirm that the series is not convergent since the general term $\displaystyle{\frac{|Q_n(x)|}{n!}}$ is not bounded.

%\clearpage

%%%NUEVA VERSION - CARLOS' MODIFICATION
\section{The Moyal star product \label{MB}}
In the non-commutative $U(1)$ theory \cite{Seiberg1999, Gomis2001_2}, we can find another intriguing example featuring operators with infinite derivatives. This theory revolves around the Moyal star product, which plays a central role. 

Given any two (smooth) functions, $f$ and $g$, the star product is defined as
\begin{equation}  \label{MB1}
f\ast g (x) = \sum_{n=0}^\infty \frac{i^{n}}{2^{n}\,n!}\,\theta^{a_1 b_1}\,\ldots \,\theta^{a_n b_n}\,\partial_{a_1 \ldots a_n} f(x)\,\partial_{b_1 \ldots b_n} g(x) \,,
\end{equation}
where $\theta^{a_j b_j}\,$ is a skewsymmetric constant matrix that we assume to be non-degenerate. 

Consider now that $f$ and $g$ are entire (real) functions on $\RR^4$, then the ID operator (\ref{MB1}) is equivalent to the following IO form
\begin{equation}\label{MB1_IO}
f \ast g (x) = \frac{1}{\pi^4 |\theta|} \int_{\RR^8} \D u\,\D v\,f(x-u) g(x-v) e^{2iv\tilde{\theta}u},
\end{equation} 
where $|\theta| = \mathrm{det}\,(\theta^{ab})$, $v\tilde{\theta}u = v^b\tilde{\theta}_{bc}u^c\,,$ and $\theta^{ab}\tilde{\theta}_{bc}=\delta^a_c$. Indeed, we can write $f$ and $g$ as formal Taylor series (\ref{i1}), namely,
\begin{equation}
f(x-u)= \sum^\infty_{n=0} \frac{(-1)^{n}}{n!} u^{a_1} \ldots u^{a_n}\partial_{a_1\ldots a_n} f(x)\,,
\end{equation}
and similarly for $g(x-v)$. Plugging them into (\ref{MB1_IO}), we arrive at
\begin{equation}\label{MB1_IO_2}
f \ast g (x) = \frac{1}{\pi^4|\theta|}\sum^\infty_{n,m=0} \frac{(-1)^{n+m}}{n!\,m!} M^{a_1\ldots a_n\,b_1\ldots b_m} \partial_{a_1\ldots a_n} f(x)\, \partial_{b_1\ldots b_m}g(x) \,,
\end{equation}
where $M^{a_1\ldots a_n\,b_1\ldots b_m}$ is 
$$ M^{a_1\ldots a_n\,b_1\ldots b_m}:=\int_{\RR^8} \D u\,\D v\,u^{a_1}\ldots u^{a_n} v^{b_1}\ldots v^{b_m}e^{2iv\tilde{\theta}u}\,,$$ and we have permuted the integrals by the infinite series because these Taylor series converge uniformly on any interval $[-K,K],\,\, K\in\RR^+$. Consequently, the series can be termwise integrated \cite{Apostol1976_V2}. The result of these integrals is
$$M^{a_1\ldots a_n\,b_1\ldots b_m}= \frac{i^n \pi^4 n!}{2^n |\tilde{\theta}|} \theta^{a_1 b_1} \ldots \theta^{a_n b_n} \delta_{nm}\,.$$
Therefore, substituting it in (\ref{MB1_IO_2}), we get\\[1ex]
%\begin{equation}
$\hspace*{4em} \displaystyle{ f \ast g (x) = \sum_{n=0}^\infty \frac{i^n}{2^n\,n!}\,\theta^{a_1 b_1}\,\ldots \,\theta^{a_n b_n}\,\partial_{a_1 \ldots a_n}f(x)\,\partial_{b_1 \ldots b_n} g(x) } \,. $\hfill $\Box$
%\end{equation}  

From now on, we shall restrict to 1+1 dimensions for the sake of simplicity, where 
$$ \theta^{a b} = \,\left( \begin{array}{cc}
												  0 & \theta \\[1ex]  -\theta & 0
													\end{array}   \right)   \,.          $$
Consider the functions $f(x^a) = f(t) \,$ and $ g(x^a) = g(x)\,$, where $\,x^a = (x,t)\,$, then the Moyal product (\ref{MB1}) reduces to
\begin{equation}  \label{MB2}
f \ast g (x,t)= \sum_{n=0}^\infty \frac{i^n \theta^n}{n! 2^n}\,(-1)^n\,f^{(n)}(t)\, g^{(n)}(x) \,.
\end{equation}

Now, we will prove that, for two particular ``harmless'' smooth functions like
\begin{equation}  \label{MB3}
 f(t) = \frac1{1+t^2}  \qquad \quad {\rm and} \qquad \quad g(x) = \frac1{1+x^2} 
\end{equation}
the series diverges. Indeed, as in Section \ref{S2.2}, write $\; f(t) = \frac12\,\left[(1+it)^{-1} + (1-it)^{-1} \right]\,$ and similarly for $g(x)\,$ to obtain 
$$ f^{(n)}(t) = \frac{n!\,i^n}{2 \,(1+t^2)^{(n+1)/2}}\,\left[e^{i \alpha (n+1)} + (-1)^n e^{-i \alpha (n+1)} \right] $$ 
and 
$$ g^{(n)}(x) = \frac{n!\,i^n}{2 \,(1+x^2)^{(n+1)/2}}\,\left[e^{i \beta (n+1)} + (-1)^n e^{-i \beta (n+1)} \right] \,, $$ 
where 
$$ \alpha = \arctan t\,, \qquad \beta = \arctan x\,, \qquad \quad \alpha, \, \beta \in \left]-\frac\pi2,\frac\pi2\right[ \,. $$ 
Substituting these derivatives into (\ref{MB2}), we have that $\,\displaystyle{ f \ast g (x,t)= \sum_{n=0}^\infty a_n(x,t)\,}$, where 
\begin{equation}  \label{MB4}
a_n (x,t) := \frac{i^n  n!\, \theta^n\, \left[\cos(\alpha+\beta)(n+1) + (-1)^n \cos(\alpha-\beta)(n+1) \right]}{2^n \left[(1+t^2)(1+x^2)\right]^{(n+1)/2}} \,.
\end{equation}
Now, a similar analysis as in Section \ref{S2.2} leads to conclude that $\,\left\{\left|a_n (x,t) \right|\,, \;\,n\in \NN\right\}\,$ is unbounded, and the series diverges for almost all values of $t$ and $x\,$.

%%%%%%%%%%%%%%%%%%%%%%%%%%%%%%%%%
\section{Conclusions}
Two approaches to the operators arise in nonlocal mechanics and nonlocal field theories, namely, one based on integral operators (IO) and another based on infinite derivatives (ID) that usually involve an infinite series. We have started by asking whether both approaches are equivalent. It is a rhetorical question because, as Moeller and Zwiebach \cite{Moeller2002} have demonstrated for the operator $\,\displaystyle{p^{\frac{1}{2}\partial^2_t} }\,$, each approach gives different results when applied to some compact support smooth function, which answers the question in the negative.

It is well-known \cite{ELIEZER1989} that both approaches can be connected through some manipulation that includes Fourier transforms. The crucial weakness of this procedure consists of the exchange of an integral and a series, which requires that some conditions are fulfilled to be legitimate.

Another point to consider is that each form of the operator, ID or IO, may act on a different functional space. We have analyzed the domains of both approaches and found that there is a common domain, but the domain of IO is broader than the one for ID. The common domain includes at least 
all entire real functions, i.e., real functions that result from restricting entire complex functions to $\mathbb{R}$. In this case, the series and the integral may be exchanged on the basis of uniform convergence of the series on compact sets. We have concentrated on the particular operator $\,\displaystyle{p^{\frac{1}{2}\partial^2_t} }\,$ and the Moyal product but the proof for a wider class of operators can be found in \cite{Carlsson2016_E4}. 

We have then examined both forms of the operator $\,\displaystyle{p^{\frac{1}{2}\partial^2_t} }\,$ and the Moyal product acting on a function that a priori seems ``harmless'' since it is smooth, bounded, and $L^p$ summable. We have seen that, although the result of applying the IO form of the operators is well-defined, the infinite series arising when applying the ID form is divergent. 

Furthermore, we have considered both forms of the operator $\,\displaystyle{p^{\frac{1}{2}\partial^2_t} }\,$ acting on a smooth compact support function, an example of the kind of functions proposed by Moeller and Zwiebach. Again, the function clearly belongs to the operator's domain in IO form, but, as we have illustrated with a numerical simulation, it does not belong in the domain of the operator's ID form because the series is not convergent. 
It is worth mentioning here the work of Carlson et al. \cite{Carlsson2016_E4} where, by using a powerful and more sophisticated mathematical formalism, it is proved that any smooth compact support function acted by the ID form of $\,\displaystyle{p^{\frac{1}{2}\partial^2_t} }\,$ yields a divergent series.

%\\{\bf (Aclarir/comentar el tipus de convergencia.)}
All this leads us to conjecture that the domain of the IO form of the operator is wider than the domain of the ID form. For this reason, it seems more advisable to use integral operators whenever possible. 

One of the reasons for using the ID form is its practicality for setting up the variational principle, Noether's theorem and a Hamiltonian formalism by merely mimicking what is done in standard local mechanics and field theory. This was the way chosen by Marnelius \cite{Marnelius1973}, who should first convert the action integral from IO form into the ID form before proceeding. However, as previously stated, the value of the proofs and derivations included in the procedure is only heuristic unless the convergence of all series is proven. For this reason, methods \cite{Heredia2021,Heredia2021_2,Heredia2022} have been recently developed to work with the same subjects, such as the variational principle, Noether's theorem, and Hamiltonian formalism, directly in the IO form.

\section*{Acknowledgment}

Funding for this work was partially provided by the Spanish MINCIU and ERDF (project ref. RTI2018-098117-B-C22) and by the Spanish MCIN (project ref. PID2021-123879OB-C22).

%% References
\bibliographystyle{utphys}
{\footnotesize\bibliography{References}}

\providecommand{\href}[2]{#2}\begingroup\raggedright\begin{thebibliography}{10}

\bibitem{Fokker1929}
A.~D. Fokker, ``Ein invarianter variationssatz f{\"u}r die bewegung mehrerer elektrischer massenteilchen,'' \href{http://dx.doi.org/10.1007/BF01340389}{{\em Zeitschrift f{\"u}r Physik} {\bfseries 58} no.~5, (1929) 386--393}.

\bibitem{Feynman1949}
J.~A. Wheeler and R.~P. Feynman, ``Classical electrodynamics in terms of direct interparticle action,'' \href{http://dx.doi.org/10.1103/RevModPhys.21.425}{{\em Rev. Mod. Phys.} {\bfseries 21} (Jul, 1949) 425--433}.

\bibitem{Pais1950}
A.~Pais and G.~E. Uhlenbeck, ``On field theories with non-localized action,'' \href{http://dx.doi.org/10.1103/PhysRev.79.145}{{\em Phys. Rev.} {\bfseries 79} (Jul, 1950) 145--165}.

\bibitem{Yukawa1950}
H.~Yukawa, ``Quantum theory of non-local fields. part i. free fields,'' \href{http://dx.doi.org/10.1103/PhysRev.77.219}{{\em Phys. Rev.} {\bfseries 77} (Jan, 1950) 219--226}.

\bibitem{Kristensen1952}
P.~Kristensen and C.~M{\o}ller {\em K Dan Vidensk Selsk Matt-Fys Medd} {\bfseries 27} (1952) 7.

\bibitem{Marnelius1973}
R.~Marnelius, ``Action principle and nonlocal field theories,'' \href{http://dx.doi.org/10.1103/PhysRevD.8.2472}{{\em Phys. Rev. D} {\bfseries 8} (Oct, 1973) 2472--2495}.

\bibitem{ELIEZER1989}
D.~Eliezer and R.~Woodard, ``The problem of nonlocality in string theory,'' \href{http://dx.doi.org/https://doi.org/10.1016/0550-3213(89)90461-6}{{\em Nuclear Physics B} {\bfseries 325} no.~2, (1989) 389--469}.

\bibitem{Biswas2006}
T.~Biswas, A.~Mazumdar, and W.~Siegel, ``Bouncing universes in string-inspired gravity,'' \href{http://dx.doi.org/10.1088/1475-7516/2006/03/009}{{\em Journal of Cosmology and Astroparticle Physics} {\bfseries 2006} no.~03, (Mar, 2006) 009--009}.

\bibitem{Modesto2012}
L.~Modesto, ``Super-renormalizable quantum gravity,'' \href{http://dx.doi.org/10.1103/PhysRevD.86.044005}{{\em Phys. Rev. D} {\bfseries 86} (Aug, 2012) 044005}.

\bibitem{Biswas2012}
T.~Biswas, E.~Gerwick, T.~Koivisto, and A.~Mazumdar, ``Towards singularity- and ghost-free theories of gravity,'' \href{http://dx.doi.org/10.1103/physrevlett.108.031101}{{\em Physical Review Letters} {\bfseries 108} no.~3, (Jan, 2012) }.

\bibitem{Frolov2016_2}
V.~P. Frolov and A.~Zelnikov, ``Radiation from an emitter in the ghost free scalar theory,'' \href{http://dx.doi.org/10.1103/physrevd.93.105048}{{\em Physical Review D} {\bfseries 93} no.~10, (May, 2016) }.

\bibitem{Deser2007}
S.~Deser and R.~P. Woodard, ``Nonlocal cosmology,'' \href{http://dx.doi.org/10.1103/physrevlett.99.111301}{{\em Physical Review Letters} {\bfseries 99} no.~11, (Sep, 2007) }.

\bibitem{Briscese2013}
F.~Briscese, A.~Marcian{\`{o} }, L.~Modesto, and E.~N. Saridakis, ``Inflation in (super-)renormalizable gravity,'' \href{http://dx.doi.org/10.1103/physrevd.87.083507}{{\em Physical Review D} {\bfseries 87} no.~8, (Apr, 2013) }.

\bibitem{Calcagni2022_2}
G.~Calcagni and L.~Modesto, ``Testing quantum gravity with primordial gravitational waves,'' \href{http://arxiv.org/abs/2206.07066}{{\ttfamily arXiv:2206.07066 [gr-qc]}}.

\bibitem{BREKKE1988}
L.~Brekke, P.~G. Freund, M.~Olson, and E.~Witten, ``Non-archimedean string dynamics,'' \href{http://dx.doi.org/https://doi.org/10.1016/0550-3213(88)90207-6}{{\em Nuclear Physics B} {\bfseries 302} no.~3, (1988) 365--402}.

\bibitem{Vladimirov2004}
V.~S. Vladimirov and Y.~I. Volovich, ``Nonlinear dynamics equation in p-adic string theory,'' \href{http://dx.doi.org/10.1023/B:TAMP.0000018447.02723.29}{{\em Theoretical and Mathematical Physics} {\bfseries 138} no.~3, (2004) 297--309}.

\bibitem{Seiberg1999}
N.~Seiberg and E.~Witten, ``String theory and noncommutative geometry,'' \href{http://dx.doi.org/10.1088/1126-6708/1999/09/032}{{\em Journal of High Energy Physics} {\bfseries 1999} no.~09, (Sep, 1999) 032--032}.

\bibitem{Gomis2001_2}
J.~Gomis, K.~Kamimura, and T.~Mateos, ``Gauge and {BRST} generators for space-time non-commutative {U}(1) theory,'' \href{http://dx.doi.org/10.1088/1126-6708/2001/03/010}{{\em Journal of High Energy Physics} {\bfseries 2001} no.~03, (Mar, 2001) 010--010}.

\bibitem{Carlsson2016_E4}
M.~Carlsson, H.~Prado, and E.~G. Reyes, ``Differential equations with infinitely many derivatives and the {B}orel transform (example 4),'' \href{http://dx.doi.org/10.1007/s00023-015-0447-4}{{\em Annales Henri Poincar{\'e}} {\bfseries 17} no.~8, (2016) 2049--2074}.

\bibitem{widder1951}
D.~Widder, ``Necessary and sufficient conditions for the representation of a function by a {W}eierstrass transform,'' {\em Transactions of the American Mathematical Society} {\bfseries 71} no.~3, (1951) 430--439.

\bibitem{Sumner1957}
D.~B. Sumner, Hirschmann, and Widder, ``The convolution transform,''
\newblock Princenton, 1957.

\bibitem{zemanian1967}
A.~H. Zemanian, ``A generalized {W}eierstrass transformation,'' {\em SIAM Journal on Applied Mathematics} {\bfseries 15} no.~4, (1967) 1088--1105.

\bibitem{Bilodeau1962}
G.~G. Bilodeau, ``{The Weierstrass transform and Hermite polynomials},'' \href{http://dx.doi.org/10.1215/S0012-7094-62-02929-0}{{\em Duke Mathematical Journal} {\bfseries 29} no.~2, (1962) 293 -- 308}.

\bibitem{Calcagni2008}
G.~Calcagni, M.~Montobbio, and G.~Nardelli, ``Localization of nonlocal theories,'' \href{http://dx.doi.org/https://doi.org/10.1016/j.physletb.2008.03.024}{{\em Physics Letters B} {\bfseries 662} no.~3, (2008) 285--289}.

\bibitem{Kolar2022}
I.~Kol{\'{a}}{\v{r}}, ``Nonlocal scalar fields in static spacetimes via heat kernels,'' \href{http://dx.doi.org/10.1103/physrevd.105.084026}{{\em Physical Review D} {\bfseries 105} no.~8, (Apr, 2022) }.

\bibitem{Moeller2002}
N.~Moeller and B.~Zwiebach, ``Dynamics with infinitely many time derivatives and rolling tachyons,'' \href{http://dx.doi.org/10.1088/1126-6708/2002/10/034}{{\em Journal of High Energy Physics} {\bfseries 2002} no.~10, (Oct, 2002) 034--034}.

\bibitem{Apostol1976_TH913}
T.~M. Apostol, {\em Mathematical Analysis (Theorem 9.13)}.
\newblock Addison-Wesley, 1960.

\bibitem{Apostol1976_V2}
T.~M. Apostol, {\em op. cit. (Theorems 9.8 and 10.27)}.

\bibitem{Apostol1976_V3}
T.~M. Apostol, {\em op. cit. (Theorem 8.11)}.

\bibitem{Apostol1976_V4}
T.~M. Apostol, {\em op. cit. (Theorem 3.24)}.

\bibitem{Vladimirov_GF_v2}
V.~S. Vladimirov, {\em Generalized functions in mathematical physics (Section: 5.2)}.
\newblock Mir, Moscow, 1979.

\bibitem{Heredia2021}
C.~Heredia and J.~Llosa, ``Energy-momentum tensor for the electromagnetic field in a dispersive medium,'' \href{http://dx.doi.org/10.1088/2399-6528/abfd14}{{\em Journal of Physics Communications} {\bfseries 5} no.~5, (May, 2021) 055003}.

\bibitem{Heredia2021_2}
C.~Heredia and J.~Llosa, ``Non-local {L}agrangian mechanics: {N}oether's theorem and {H}amiltonian formalism,'' \href{http://dx.doi.org/10.1088/1751-8121/ac265c}{{\em Journal of Physics A: Mathematical and Theoretical} {\bfseries 54} no.~42, (Sep, 2021) 425202}.

\bibitem{Heredia2022}
C.~Heredia and J.~Llosa, ``Nonlocal lagrangian fields: Noether's theorem and hamiltonian formalism,'' \href{http://dx.doi.org/10.1103/physrevd.105.126002}{{\em Physical Review D} {\bfseries 105} no.~12, (Jun, 2022) }.

\end{thebibliography}\endgroup

\end{document}